\documentclass[aps,prl,amsmath,twocolumn,amssymb,titlepage,showpacs]{revtex4-1}

\usepackage{graphicx}
\usepackage{nicefrac}
\usepackage{amsfonts}

\usepackage{amssymb}
\usepackage{amsmath} 

\usepackage{subfigure}
\usepackage{multirow} 
\usepackage{tabularx} 
\usepackage{array}

\usepackage{units}

\usepackage{tensor} 
\usepackage{braket}

\usepackage{bm}
\usepackage{hyperref}

\newcommand{\tr}{\operatorname{{\mathrm tr}}}

\newcommand{\inbk}[1]{\left[ #1 \right]}

\begin{document}

\title{Probing the structure of entanglement with entanglement moments}
\author{Justin H.\ Wilson}
\affiliation{Joint Quantum Institute and Condensed Matter Theory
  Center, Department of Physics, University of Maryland, College Park,
  Maryland 20742-4111, USA} 
\author{Joe Mitchell}
\affiliation{Joint Quantum Institute and Condensed Matter Theory
  Center, Department of Physics, University of Maryland, College Park,
  Maryland 20742-4111, USA} 
\author{Victor Galitski}
\affiliation{Joint Quantum Institute and Condensed Matter Theory
  Center, Department of Physics, University of Maryland, College Park,
  Maryland 20742-4111, USA} 


\begin{abstract}
  We introduce and define a set of functions on pure bipartite states called \emph{entanglement moments}. 
  Usual entanglement measures tell you \emph{if} two systems are entangled, while entanglement moments tell you both \emph{if} and \emph{how} two systems are entangled.
  They are defined with respect to a measurement basis in one system (e.g., a measuring device), and output numbers describing how a system (e.g., a qubit) is entangled with that measurement basis. 
  The moments utilize different distance measures on the Hilbert space of the measured system, and can be generalized to any $N$-dimensional Hilbert space. 
  As an application, they can distinguish between projective and non-projective measurements.
  As a particular example, we take the Rabi model's eigenstates and calculate the entanglement moments as well as the full distribution of entanglement.
\end{abstract}


\maketitle

Quantifying entanglement has been of interest since Bell showed that this uniquely quantum feature was available for experimental verification \cite{Bell1964, *Bell1966, *Aspect1981}. 
Since Bell, we have seen an explosion of potential applications in quantum information \cite{Ekert1991} and computation \cite{Horodecki2009} as well as a whole body of theory to address the quantification of entanglement \cite{Plenio2007}.
There are many measures of entanglement for pure states and mixed state \cite{Plenio2007} which take as input a state and outputs a number telling you, very roughly speaking, how entangled a state is. 
However, these measures just tell one \emph{if} a state is entangled, but not \emph{how} it is entangled: two very different states can give the same number.
To address this, we define a new set of functions called \emph{entanglement moments}.
(While we call these entanglement ``moments'', they are not moments in the usual sense of distributions.)
These quantities can tell us not only if and by how much a state is entangled but also how the distribution of entanglement looks by telling us how ``clumpy'' our distribution is. 

\begin{figure}
  \includegraphics[width=0.45\textwidth]{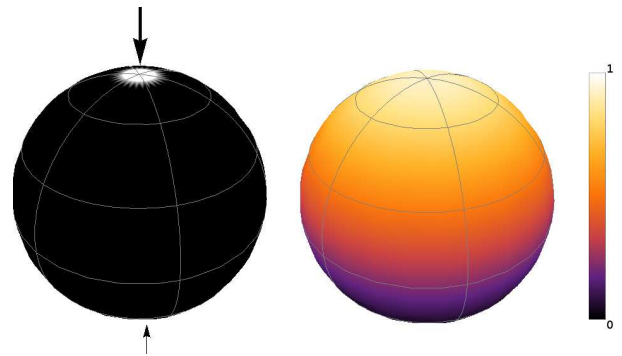}
  \caption{(Color online) Shown are two different distributions of entanglement on the Bloch sphere.
  The left consists of two delta functions, on the north and south poles, with possibly different magnitudes (the arrow sizes are proportional to the strength of the delta functions).
  The right is a smoother distribution.
  The two have the same concurrence, yet are qualitatively different distributions of entanglement.
  Entanglement moments can distinguish them.}
  \label{fig:motivation}
\end{figure}

For example, if we have a qubit entangled with another system and we make measurements on that other system, we will get a distribution of qubit states on the Bloch sphere.
Two such examples are shown in Fig.~\ref{fig:motivation}. 
Note that in the first case, the distribution is centered around the north and south pole; while in the second case, the distribution is more evenly distributed about the sphere.
We would like a measure that can distinguish these two instances -- both of which have the same entanglement as given by the usual entanglement measures such as concurrence \cite{Wootters1998,Rungta2001}.


As an application, the property of entanglement moments to describe how the system is entangled allows them to characterize measurements from weak to strong/projective measurements \cite{Zurek1981,*Zurek1982}.
This uses the prescription for quantum measurement where the apparatus is treated quantum mechanically, becoming entangled with the system and mediating the collapse of the wave function \cite{VonNeumann1955}. 
This phenomenon has been exploited to understand the measurement process in the lab in terms of finite strength quantum measurement (see for instance \cite{Brune1996,*Vijay2012,*Hatridge2013}).
In this situation, a measuring device is entangled with another system, and making measurements on the device indirectly probes the second system in what may be a non-projective way.
Considering the apparatus and system, a projective measurement corresponds to a very ``clumpy'' distribution in the system's Hilbert space while a non-projective measurement would be more evenly distributed. 
The entanglement moments can tell the difference between these two distributions and hence between projective and certain non-projective measurements.

In this letter, we first define entanglement moments for two systems $A$ and $B$ where $A$ is the measured system.
Then, without specifying the system $B$ doing the measurement, we analyze the entanglement moments when our measured system $A$ is a qubit restricted to $S^1$ on the Bloch sphere and when $A$ is the entire Bloch sphere.
The analysis generalizes to when $A$ is an arbitrary $N$-dimensional Hilbert space in addition to the qubit case (see section 1 of the supplement \cite{supplement}).
We then analyze the expressions with some informative examples. 

\begin{figure}
  \includegraphics[width=0.45\textwidth]{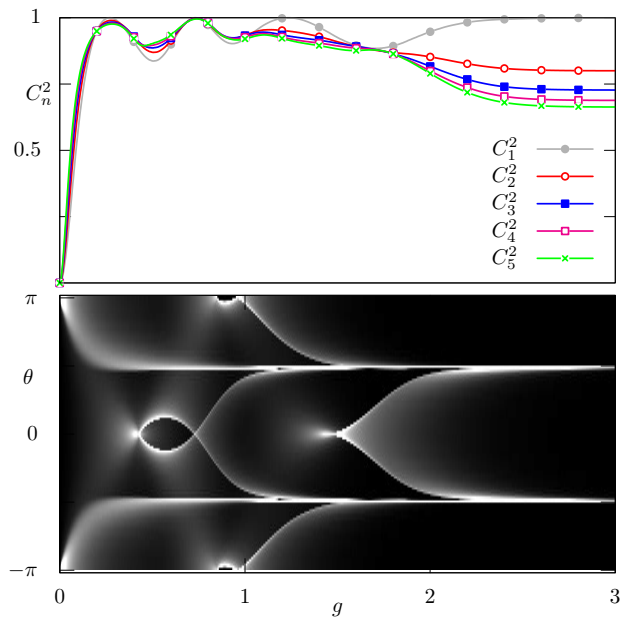}
  \caption{(Color online) The sixth excited state of the Rabi model with $\Delta=0.3$.  The concurrence, $C_{(1)}^2$, does not hold the information necessary to deal with such a complicated state.  Top:  Entanglement moments.  Bottom:  Plot of the distribution along the relevant circle on the Bloch sphere. Notice how the entanglement is non-monotic with respect to the coupling $g$.}
  \label{fig:Rabi6Excited0.3}
\end{figure}

Finally, to illustrate a physically relevant application of entanglement moments, we analyze the Rabi model \cite{Rabi1936}. 
This model shows up in many areas of physics including but not limited to circuit QED \cite{Niemczyk2010}, cavity QED \cite{Schiro2012, Englund2007}, photonics \cite{Crespi2012}, and flux qubits \cite{Forn-Diaz2010}.
It comes into play when a qubit and a harmonic oscillator interact, and hence it finds its way into many of the approaches to quantum computation \cite{Pellizzari1995}. 
While entanglement in the Rabi model has been studied before \cite{Fang1996}, we show how the entanglement moments track the distribution of qubit states in particular eigenstates of the system -- demonstrating how the moments discriminate between projective and non-projective measurements. 
The results are obtained numerically from the exact solution recently found by Braak \cite{Braak2011}. 
In addition, the distribution of qubit states shows one way in which the Rabi model can give qualitatively different results from the Jaynes-Cummings model such as the non-monotonic behavior of entanglement with respect to the interaction strength (seen in Fig.~\ref{fig:Rabi6Excited0.3}).

In order to address this question of ``\emph{how}'' two systems can be entangled, we lift one of the requirements of entanglement measures: that the measure must be invariant under local unitary transformations \cite{Plenio2007}.  
To understand why we need to lift this requirement in order to get at the nature of the entanglement, consider a qubit $A$ coupled to any system $B$ with $\ket\Psi = \ket \uparrow \otimes \ket{\phi_1} + \ket \downarrow \otimes \ket{\phi_2}$.
Given any basis for $B$, $\{\ket 1, \ket 2, \ldots, \ket N\}$, we can always perform a unitary operation, $U$, on the Hilbert space of $B$ such that $U\ket{\phi_1} = a\ket 1$ and $U \ket{\phi_2} = b \ket 1 + c \ket 2$. 
This unitary operation has destroyed the information describing \emph{how} the basis is entangled with our qubit. 
While it is true that only two states at any given time are entangled with a qubit, rarely can an experiment know what those two states are \emph{a priori}. 
On the other hand, we must take into account the converse of this: 
We can rotate $a\ket 1$ and $b\ket 1 + c \ket 2$ into any two vectors (such that $\braket{\phi_1|\phi_2} = a^* b$). 
Thus, the choosing of a basis must have some physical relevance and hence we call it the \emph{measurement basis}.

To construct these entanglement moments, consider systems $A$ and $B$ and let $\{\ket 1, \ket 2, \ldots, \ket N\}$ be the measurement basis in system $B$.
We can write a general state vector as
\begin{align}
\label{eq:state-vector}
    \ket\Psi = \sum_i \ket{\psi_i} \otimes \ket i,
\end{align}
where $\ket{\psi_i}$ are unnormalized vectors in the Hilbert space of $A$. Defining the weight of the vector by $|\psi_i|^2 \equiv \braket{\psi_i|\psi_i}$, we write the expression for $n$th entanglement moment as
\begin{align}
\label{eq:entangle-moment}
    C^2_{(n)} = \mathcal N_n \sum_{i,j} |\psi_i|^2 d^2_{(2n)}(i,j) |\psi_j|^2,
\end{align}
where for the normalized $\ket{\tilde \psi_i} = \ket{\psi_i}/|\psi_i|$,
\begin{align}
\label{eq:distance-measure}
	d^2_{(2n)}(i,j) = 1 - | \braket{\tilde \psi_i|\tilde\psi_j}|^{2n},
\end{align}
and the quantity $\mathcal N_n$ normalizes the maximal value of $C^2_{(n)}$ to unity -- we will specify its value for specific cases later.
The quantity $d^2_{(2n)}(i,j)$ is a distance function on the Hilbert space of system $A$. 
If $n=1$, we get the Hilbert-Schmidt distance measure (which leads to the Fubini-Study metric), and if we let $n \rightarrow \infty$, we get the trivial distance measure.
The presence of the distance measure is to quantify how system $A$ changes as we measure system $B$; if system $A$ changes upon measurement of system $B$, we know they are entangled.
Therefore, a state $\ket\Psi$ is separable if and only if each entanglement moment is zero.

For the specific value $n=1$, we actually reproduce I-Concurrence \cite{Rungta2001} in general (or just concurrence in the case of a qubit):
\begin{align}
\label{eq:concurrence}
 C^2_{(1)} 
   & = \mathcal N_1 \inbk{1 - \tr \varrho_A^2},
\end{align}
where $\varrho_A$ is the reduced density matrix of system $A$ and $\tr \varrho_A^2$ is the order-2 R\'enyi entropy \cite{Bengtsson2006}. 
Thus, the quantity $C^2_{(1)}$ is invariant under all local unitary transformation.
Not only is this quantity I-Concurrence, but it begins with a clearer, geometric, and intuitive definition (Eq.~(\ref{eq:entangle-moment})).

To illustrate what happens when $n>1$, we first assume that not only is system $A$ on the Bloch sphere but that our states are constrained to be on the great circle $S^1$ defined by the $y$-axis. 
(This example is illustrative: we consider the entire Hilbert space in Eq.~(\ref{eq:sphere-entanglement-char}).)
We can define the distribution of the entanglement $\rho : S^1 \longrightarrow \mathbb R$ such that $\rho(\theta)$ is the probability distribution of states in the Hilbert space of $A$ given measurements in $B$ with a particular basis (each $\theta$ corresponds to a particular state in Hilbert space). 
This allows us to rewrite the entanglement moments as
\begin{align}
\label{eq:circle-entanglement-char}
	C^2_{(n)} = \mathcal N_n \int_{-\pi}^{\pi} d \theta\, d\theta' \, \rho(\theta) d^2_{(2n)}(\theta,\theta')\rho(\theta').
\end{align}
(In the case of a countable number of $\ket{\psi_i}$'s, $\rho(\theta)$ is a sum of delta functions.) In this representation, we know the exact form of the distance measure $d^2_{(2n)}(\theta,\theta') = 1 - \cos^{2n}[(\theta-\theta')/2]$.
With simple trigonometric identities, it can be shown that
\begin{align}
\label{eq:distance-expand-S1}
	d^2_{(2n)}(\theta,\theta') 
	  & = 1 - \frac1{2^{2n}}\binom{2n}{n} - \frac2{2^{2n}} \sum_{k=1}^n \binom{2n}{n-k} e^{ik(\theta-\theta')}.
\end{align}
Substituting this into Eq.~(\ref{eq:circle-entanglement-char}) and using the fact that we have normalized $\rho(\theta)$ such that $\int_{-\pi}^{\pi}d\theta \rho(\theta)=1 $, we obtain
\begin{multline}
\label{eq:circle-entanglement-char-2}
	C^2_{(n)} = \mathcal N_n \left[ 1 - \frac1{2^{2n}}\binom{2n}{n} \right. \\ \left. - \frac2{2^{2n}} \sum_{k=1}^n \binom{2n}{n-k}\left| \int_{-\pi}^{\pi} d\theta \rho(\theta) e^{i k \theta}\right|^2 \right].
\end{multline}
We can read off the normalization
\begin{align}
\label{eq:normalization-S1}
 \mathcal N_n[S^1] = \frac{2^{2n}}{2^{2n} - \binom{2n}n}.
\end{align}

These entanglement moments are picking up the features of this distribution in terms of its Fourier components -- the distance functions are diagonal in this basis.

This expression gives us insight into our original expression for entanglement moments. 
First, the $n$th moment decreases from unity for each Fourier coefficient up to the $n$th that is non-zero. 
Corollary to this, since $C^2_{(1)}$ is invariant under local unitary transformations, the norm of the first Fourier component will remain the same no matter what measurement basis is chosen (or in the general case, the occupation in the first harmonic will remain the same; see section 1 of the supplement \cite{supplement}). 
By the uncertainty principle, highly localized (separable) states have a large distribution in frequency space, so such states will have large numbers of suppressed entanglement moments. 

Consider the equation for entanglement moments Eq.~(\ref{eq:entangle-moment}).
As we increase $n$, the distance measures interpolate between points on the circle to the trivial distance measure (every point is a distance $1$ away from every other point). 
If we now consider a state localized near the north and south poles,
the points near the north pole are roughly a distance $1$ from all points near the south pole for all distance measures $d^2_{(2n)}$. 
Thus, each $C^2_{(n)}$ decreases solely because the normalization $\mathcal N_n$ decreases. 
In this way, a decreasing of the moments is indicative of there being more than one ``clump'' in the entanglement distribution.
Hence, if measuring system $B$ corresponds to a projective measurement on $A$, one should see the entanglement moments all decrease.

\begin{figure}
  \includegraphics[width=0.45\textwidth]{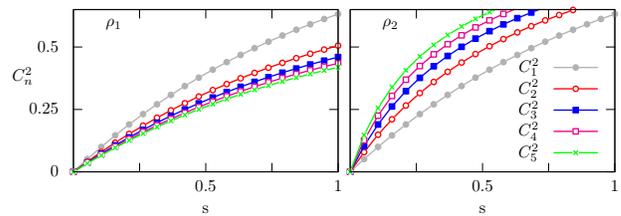}
  \caption{(Color online) Entanglement moments for distributions $\rho_1$ and $\rho_2$ from Eq.~(\ref{eq:two-distributions}) on the circle $S^1$. $\rho_1$ changes from localized to the north pole to equally localized to north and south, while $\rho_2$ delocalizes from the north pole to cover the entire circle as $s\to\infty$.  As $s\to\infty$, the moments of $\rho_2$ will limit to $1$, while those of $\rho_1$ will be less.}
  \label{fig:MotivatingMoments}
\end{figure}

For further illustration, consider two distributions on $S^1$:
\begin{align}
\label{eq:two-distributions}
  \rho_1(\theta) & = \frac12\left[ (1+e^{-s}) \delta(\theta) + (1- e^{-s}) \delta(\theta - \pi) \right] \\
\notag
  \rho_2(\theta) & = \frac1{2\pi} \sum_{k=-\infty}^\infty e^{-k^2 s} e^{ik\theta},
\end{align}
where $s$ is an arbitrary parameter, and $\rho_1(\theta; s=0) = \rho_2(\theta; s=0) = \delta(\theta)$.
These distributions are chosen such that they normalize to unity, and for every $s$, they give the same entanglement $C^2_{(1)}[\rho_1] = C^2_{(1)}[\rho_2]$.
While on $S^1$ rather than the entire Bloch sphere, they are qualitatively like the two distributions in Fig.~\ref{fig:motivation}.
The first is a projective measurement, interpolating between the unentangled state at $s=0$ and equal probable projection as $s\to\infty$. 
The second is a state localized around $\theta=0$, delocalizing as $s$ increases to eventually cover the whole circle (it is in fact the Green's function of the heat equation).  
Plotting the higher entanglement moments in Fig.~\ref{fig:MotivatingMoments}, we see a stark contrast. 
In the first all moments get smaller as we increase $n$, and in the second they get larger.
While we have explained this in terms of Eq.~(\ref{eq:entangle-moment}), consider now the mode decomposition Eq.~(\ref{eq:circle-entanglement-char-2}): the higher modes in $\rho_2(\theta)$ are exponentially suppressed; while in $\rho_1(\theta)$, half of the higher modes stay constant.

\begin{figure}
  \includegraphics[width=0.45\textwidth]{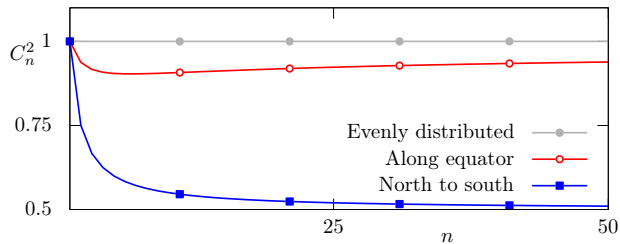}
  \caption{(Color online) Entanglement moments for distributions on the Bloch sphere.  All have maximal concurrence $C_{(1)}^2=1$, but the moments of the more localized distributions deviate greatly.  The $n\to\infty$ limit of the distribution localized to north and south poles reveals its pointlike nature.}
  \label{fig:EntVsN}
\end{figure}

If we know our system is highly entangled, we can use the moments to tell if measuring system $B$ will correspond to a projective measurement on system $A$. 
To illustrate this, we step from $S^1$ to the entire Bloch sphere $S^2$. 
The results are
\begin{align}
\label{eq:sphere-entanglement-char}
	C^2_{(n)} = \frac{n+1}{n} \int_{S^2} d \mathbf{n}\, d \mathbf{n}' \, \rho(\mathbf{n}) d^2_{(2n)}(\mathbf{n},\mathbf{n}')\rho(\mathbf{n}'),
\end{align}
\begin{multline}
\label{eq:sphere-entanglement-char-expand}
	C^2_{(n)} = 1 - \frac{n+1}n \sum_{l=1}^n \frac{\binom n l}{\binom{n+l+1}n} \frac{4\pi}{l+1} \\ \times \sum_{m=-l}^l \left|\int_{S^2} d \mathbf{n} \, \rho(\mathbf{n}) Y_{lm}^*(\mathbf{n}) \right|^2 .
\end{multline}
(The generalization to an $N$-dimensional Hilbert space is found in section 1 the supplement \cite{supplement}.) 
We see that instead of Fourier coefficients, we now deal with harmonics: $ \sum_{m} \left|\int_{S^2}  \rho  Y_{lm}^* \right|^2$.
To showcase this analysis on the sphere, Fig.~\ref{fig:EntVsN} shows the entanglement moments for three distributions that all have the same, maximal entanglement as measured by concurrence (i.e.\ $C_{(1)}^2=1$): evenly distributed about the sphere, evenly localized along the equator, and localized to the north and south poles. 
Note how in each case, the higher entanglement moments indicate how localized our state is. 
The highly localized state at the north and south pole asymptote down to $\frac12$, indicating it is in fact localized to two points. 
For the distribution localized to the equator, it dips down, showing that it is localized, but rises back up to asymptote to 1. 
The reason it rises is that no state in its distribution is localized to a set of measure zero -- if we go to our original expression for entanglement and let $n\rightarrow \infty$, then $d_{(\infty)}^2 (i,j)= 1$  if $\ket{\tilde \psi_i}$ and $\ket{\tilde\psi_j}$ are not the same and it is 1 if they are. 
So if we have such an even distribution of states, $C^2_{(\infty)} = 1$ just by the normalization of our states: $\braket{\Psi|\Psi}=1$.

For a more in depth example, we consider a qubit and a harmonic oscillator described by the Rabi Hamiltonian
\begin{align}
\label{eq:Rabi-model}
	H_{\text{Rabi}} = \omega a ^{\dagger} a + g \sigma_x (a + a^\dagger) + \tfrac12 \Delta \sigma_z,
\end{align}
where $a$($a^\dagger$) is the annihilation (creation) operator for the harmonic oscillator and $\sigma_i$ are the Pauli matrices for the qubit.

If we specify our measurement basis as the eigenbasis for the operator $\hat x= \frac1{\sqrt2}(a + a^\dagger)$, we can write a vector in this Hilbert space as
\begin{align}
\label{eq:vector}
  \ket{\psi} = \int dx \ket{\psi(x)} \otimes \ket{x},
\end{align}
where $\ket{\psi(x)}$ is a vector on the Bloch sphere. 
This set of vectors can be mapped onto a distribution $\rho$ on the Bloch sphere.

\begin{figure}  \includegraphics[width=0.45\textwidth]{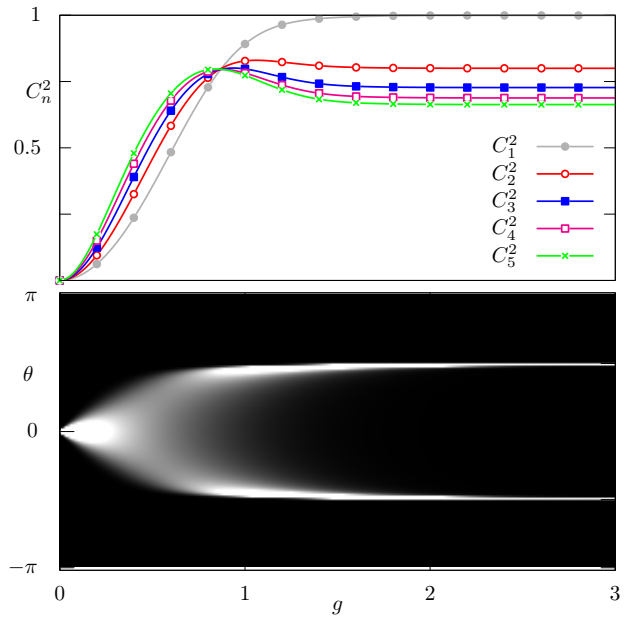}
  \caption{(Color online) Rabi model ground state with $\Delta=0.3$.  Top:  Entanglement moments.  Note that the crossover corresponds roughly with the distribution forming into two nonoverlapping localized points.  Bottom:  Plot of the distribution along the relevant circle on the Bloch sphere.}
  \label{fig:RabiGround0.3}
\end{figure}

Now consider explicitly the eigenstates of Eq.~(\ref{eq:Rabi-model}). 
They can be labeled by an integer and $\pm$ as shown in the exact solution given by Braak \cite{Braak2011}.
These states, $\ket{n,\pm}$, only live on a circle $S^1$ of the Bloch sphere due to the exclusion of $\sigma_y$ from the Hamiltonian. 
As such, we use Eq.~(\ref{eq:circle-entanglement-char}) for the entanglement moments (for details, see section 2 of the supplement \cite{supplement}). 
The entanglement moments and full distributions for $\Delta=0.3$ are plotted for the ground state in Fig.~\ref{fig:RabiGround0.3} and for the sixth excited state in Fig.~\ref{fig:Rabi6Excited0.3}.

Notice in these figures how the distribution changes with $g$ for a given eigenstate. 
For large $g$, the higher moments asymptote to a value less than 1 while $C_{(1)}^2 \rightarrow 1$. 
This represents the localization described in Fig.~\ref{fig:EntVsN} as well as how a measurement in $\hat x$ corresponds to a projective measurement in $\sigma_x$ of the qubit.
On the other hand, the figures show other instances where $C_{(1)}^2 = 1$ along with other entanglement moments, and in those cases the states are more evenly distributed about the circle.
In fact when we see moments rise for higher values of $n$, we know the state is becoming more evenly distributed just as in the case of Fig.~\ref{fig:MotivatingMoments}.
The cross-over from the moments getting larger as $n$ increases to the point where they start decreasing with $n$ represents the cross-over from non-projective to projective-like measurements.

Entanglement moments could also be used in various dynamical questions and in many other systems (such as the system considered in \cite{wilson2012entanglement}).
Additionally, generalization to density matrices remains an open question at this time.
Such a theory would need to generalize I-concurrence and separate out the classical probabilities from the quantum.

In this letter, we have defined the new concept of entanglement moments.  
These moments contain and surpass traditional entanglement measures, describing not only if a system is entangled, but also how.  
Taken all together, they can qualitatively and quantitatively describe how projective a measurement is.  
As seen in Figs.~\ref{fig:MotivatingMoments} and~\ref{fig:EntVsN}, if higher moments increase, the distribution will be quite distributed and the measurement less projective.  
As an example, we calculated these moments for eigenstates of the Rabi model, showing complex behavior for higher excited states.




This research was supported by DOE-BES-DESC0001911 (V.G. and J.M.) and the JQI-PFC  (J.W.)

\bibliography{references}

\onecolumngrid

\hspace{12pt}

\hrule

\hspace{12pt}

\section{Supplementary material: Probing the structure of entanglement with entanglement moments}

\twocolumngrid

In this supplement, we discuss entanglement moments when the measured system is an $N$ dimensional Hilbert space, and we detail the Rabi model calculations presented in the main text.
We described how to apply entanglement moments when the measured system is $S^1$ and the Bloch sphere, but the analysis is far more general.
By utilizing the mathematics of $\mathbb C \mathrm{P}^{N-1}$, we can apply entanglement moments in $N$ dimensions.
For completeness, we also discuss our calculations in the Rabi model in depth.
A wealth of information in the Rabi model is obtainable with exact calculations and simple numerics.

\section{Extending analysis to any finite dimensional Hilbert space}

We assume a bipartite system where the Hilbert space is the direct product of two other Hilbert spaces $\mathcal H = \mathcal H_A \otimes \mathcal H_B$. 
We can write states $\ket{\Psi}\in \mathcal H$ in terms of an orthonormal basis of $\mathcal H_B$, $\{ \ket 1, \ket 2, \ldots, \ket{N_B}\}$, and (unnormalized) vectors $\ket{\psi_i}\in\mathcal H_A$,
\begin{align}
\label{eq:psi}
  \ket{\Psi} = \sum_i \ket{\psi_i} \otimes \ket{i}.
\end{align}
The vector $\braket{\Psi|\Psi}=1$ while $\braket{\psi_i|\psi_i}\leq 1$ in general.

Let system $\mathcal H_A$ be an arbitrary $N$-dimensional Hilbert space. 
The space of normalized vectors is $S^{2N-1}$, but there is a U(1) gauge freedom in the distance measures given by 
\begin{align}
\label{eq:distance-measure-sup}
  d^2_{(2n)}(i,j) = 1 - |\braket{\tilde\psi_i|\tilde\psi_j}|^2,
\end{align}
so the space is actually $\mathbb C \mathrm{P}^{N-1} = S^{2N-1}/\text{U}(1)$. 
This is in fact a Hopf fibration from $S^{2N-1}$ to $\mathbb C \mathrm P^{N-1}$ over the U($1$) fiber. 
As with the other cases considered in the main text, we map $\braket{\psi_i|\psi_i}$ onto the function $ \rho:\mathbb{C}\mathrm{P}^{N-1}\longrightarrow \mathbb{R}^+$. 

The entanglement moments are then given by
\begin{align}
\label{eq:entanglement-moment-CP}
    C^2_{(n)} = \mathcal N_n \int_{\mathbb C \mathrm{P}^{N-1}}\hspace{-20pt}d\mu(z)\int_{\mathbb C \mathrm{P}^{N-1}} \hspace{-20pt} d\mu(w) \, \rho(z) d^2_{(2n)}(z,w) \rho(w).
\end{align}

As in the main text, the distance functions are known:
\begin{align}
\label{eq:complex-projective-distance}
    d^2_{(2n)}(z,w) = 1 - \Big|\sum_i z_i^* w_i\Big|^{2n}.
\end{align}
The distance function can be considered as a function of the on the \emph{angle set}, so that we write
\begin{align}
\label{eq:angle-set-expr}
  d^2_{(2n)}(z,w) = d^2_{(2n)}(2\Big|\sum_i z_i^* w_i\Big|^2 - 1).
\end{align}
Since Eq.~(\ref{eq:angle-set-expr}) is an $n$th ordered polynomial in $2\left|\sum_i z_i^* w_i\right|^2 - 1$, we can expand it into Jacobi polynomials, $P_k^{(N-2,0)}(2\left|\sum_i z_i^* w_i\right|^2 - 1)$. 

Now we need to use an addition formula for complex projective space as derived by \cite{Koornwinder1972sup,*Koornwinder1972asup,Shatalov2001sup}.
To develop the formula, we should write the space of functions, $L^2(\mathbb{C}\mathrm{P}^{N-1})$, as a direct sum of orthogonal subspaces in the following way.
Dividing into the spaces of spherical harmonics, we have $L^2(S^{2N-1}) = H_1(2N) \oplus H_2(2N) \oplus \cdots$, where $H_m(2N)$ is the finite-dimensional vector space of harmonic polynomials homogeneous of degree $m$ of $2N$ real variables that are restricted to $S^{2N-1}$. 
These should be further restricted to those that are just U($1$) invariant since $\mathbb{C}\mathrm{P}^{N-1} = S^{2N-1}/\mathrm{U}(1)$.
With this restriction, we follow the notation of \cite{Grinberg1983asup} and write
\begin{align}
\label{eq:ltwocpn-decomp}
    L^2(\mathbb{C}\mathrm{P}^{N-1}) = H_{(0,0)}(N) \oplus H_{(1,1)}(N)  \oplus H_{(2,2)}(N) \oplus \cdots,
\end{align}
where $H_{(m,m)}(N)$ are just the U($1$) invariant parts of $H_m(2N)$.

Given this, we now state the addition theorem as written in \cite{Shatalov2001sup}. 
Let $d_{k,N} = \dim H_{(k,k)}(N)$ and $s_{kj}$ be an orthonormal basis in the space $H_{(k,k)}(N)$.
Then the Jacobi polynomials become
\begin{multline}
\label{eq:addition-theorem-general}
    P_k^{(N-2,0)}\left(2\Big|\sum_i z_i^* w_i\Big|^2 - 1\right) \\ = \frac1{d_{k,N}}\binom{k+N-2}{k} 
  \sum_{j=1}^{d_{k,N}} s^*_{kj}(z) s_{kj}(w).
\end{multline}

Note that we can also calculate $d_{k,N}$ from formulae given in \cite{Shatalov2001sup}. 
It is
\begin{align}
\label{eq:dimension-of-Hkk}
  d_{k,N} = \frac{2k+N-1}{N-1} \binom{k+N-2}{k}^2.
\end{align}
Just as before, we can expand our the distance function in terms of $P_k^{(N-2,0)}(2\left|\sum_i z_i^* w_i\right|^2 - 1)$, then expand that by the addition theorem and obtain
\begin{align}
\label{eq:entanglement-moments}
  C^2_{(n)} = \mathcal N_n \inbk{1 - \sum_{k=0}^n \frac{\binom{n}k}{\binom{k+n+N-1}{n}\binom{k+N-1}{k}} \lVert \rho \rVert^2_{H_{(k,k)}(N)}},
\end{align}
where 
\begin{align}
\label{eq:norm-in-Hkk}
  \lVert \rho \rVert^2_{H_{(k,k)(N)}} = \sum_{j=1}^{d_{k,N}} \left| \int_{\mathbb C \mathrm{P}^{N-1}}\hspace{-20pt}d\mu(z) \, \rho(z) s_{kj}^*(z) \right|^2
\end{align}
is the norm of the function in the finite subspace $H_{(k,k)}(N)$ -- i.e., the norm in the $k$th harmonic. 
So the $n$th entanglement moment captures the information about the 1st through $n$th harmonic of the distribution.

Proper normalization of our distribution gives us $\lVert \rho \rVert^2_{H_{(0,0)}(N)}=1$, since $H_{(0,0)}(N)$ is the space of constant functions.  We can read off the normalization as
\begin{align}
\label{eq:normalization}
  \mathcal N_n[\mathbb C \mathrm P ^{N-1}] = \frac{\binom{n+N-1}{n}}{\binom{n+N-1}{n}-1}.
\end{align}

This entire analysis reduces to the case of a Bloch sphere for $N=2$, and we reproduce the Bloch sphere formula from the main text exactly.

\section{Entanglement calculations in the Rabi model}

For the Rabi model, $\mathcal H_A$ is a two level system and $\mathcal H_B$ is a harmonic oscillator, and we have
\begin{align}
\label{eq:hamiltonian-rabi}
  H = \omega a^\dagger a + g (a + a^\dagger) \sigma_x + \Delta \sigma_z,
\end{align}
where $a$($a^\dagger$) is the annihilation (creation) operater, $\sigma_x$ and $\sigma_z$ are the $x$ and $z$ Pauli matrices respectively, and $\omega$, $g$, and $\Delta$ are constants (frequency of the oscillator, coupling, and Zeeman splitting, respectively). 

The Rabi model's eigenstates have a particular form since the operator $\sigma_z\otimes P$, where $P$ is the parity operator on the harmonic oscillator, commutes with the Hamiltonian. The states are
\begin{align}
\label{eq:eigenstates}
  \ket{\Psi_\pm} = \frac1{\sqrt2}\left[ \ket{+} \otimes \ket{\phi_\pm} \pm \ket{-} \otimes P\ket{\phi_\pm}\right]
\end{align}
where $\ket +$ and $\ket -$ are the eigenstates of $\sigma_x$ with eigenvalues $\sigma_x\ket{\pm} = \pm \ket{\pm}$, and $\ket{\phi_\pm}$ are unknown vectors in the Hilbert space of the harmonic oscillator. 

The concurrence \cite{Rungta2001sup} for this system can be easily calculated, and happens to be
\begin{align}
\label{eq:concurrence-rabi}
  C^2_{(1)} = 1 - |\braket{\phi_\pm| P | \phi_\pm}|^2,
\end{align}
so the entanglement of these states just depends on the expectation value of the parity operator with  harmonic oscillator wave functions.

We can exactly calculate things if we let $\Delta \rightarrow 0$. In that case the eigenstates are just
\begin{align}
\label{eq:eigenstates-with-P}
  \ket{\Psi_\pm; \Delta\rightarrow 0} = \frac1{\sqrt2}\left[ \ket{+} \otimes \ket{n}_L \pm \ket{-} \otimes P\ket{n}_L\right],
\end{align}
where $\ket{n}_L$ is the $n$th state of the harmonic oscillator centered at $x=-\sqrt 2 g$.

The resulting concurrence is then
\begin{align}
\label{eq:concurrence-Delta-zero}
  C^2_{(1)} = 1 - L_n(g^2/2)e^{-g^2/4},
\end{align}
where $L_n(g^2/2)$ are Lagueere polynomials. As the coupling $g$ increases, we see oscillations in the entanglement due to the Laguerre polynomials.

Braak \cite{Braak2011sup} solved for the eigenstates when $\Delta \neq 0$.
The eigenvalues can be calculated from $E_n^\pm = \xi_n^\pm-g^2/\omega$ and
\begin{align}
\label{eq:eigenvalue equation}
  0 = G_\pm(\xi_n^\pm) = \sum_{m=0}^\infty K_m(\xi_m^\pm) \left[1 \mp \frac{\Delta}{\xi_m^\pm - m \omega}\right] \left( \frac g \omega \right)^2,
\end{align}
where the coefficients $K_m(\xi)$ satisfy
\begin{align}
\label{eq:coefficients}
  m K_m &= f_{n-1}(\xi) K_{m-1} - K_{m-2},\\ K_0=1, &\quad K_1(\xi) = f_0(\xi),
\end{align}
\begin{align}
\label{eq:ffunction}
  f_m(\xi) = \frac{2g}\omega + \frac1{2g} \left(m \omega - \xi + \frac{\Delta^2}{\xi - m \omega} \right).
\end{align}
The unnormalized eigenstates, written in Bargmann space \cite{Bargmann1961sup}, are
\begin{multline}
\label{eq:eigenstatessup}
  \phi^\pm_n(z) = e^{g z } \sum_{n=0}^\infty K_n(\xi_n^\pm) (-z + g)^n \\ = \pm e^{g z } \sum_{n=0}^\infty K_n(\xi_n^\pm) \Delta \frac{(z + g)^n}{\xi_n^\pm- n}.
\end{multline}
Taking normalization into account, we can obtain
\begin{align}
\label{eq:concurrencesup}
  C^2_{(1)} [ \ket{\Psi_n^\pm}] =  1 - \left( \frac{\sum_{n=0}^\infty n! K_n(\xi_n^\pm)^2  \frac \Delta{\xi_n^\pm - n}}{ \sum_{n=0}^\infty n! K_n(\xi_n^\pm)^2 }\right)^2.
\end{align}


The higher moments, $C^2_{(n)}$, do not admit a closed form expression when we specify the measurement basis as the eigenstates of $\hat x = \frac1{\sqrt2} (a + a^\dagger)$.
However, it is straightforward to numerically calculate them.
The eigenvalues can be calculated from Eq. \ref{eq:eigenvalue equation} using a relatively simple rootfinding algorithm.
Then with the eigenfunctions $\left( \psi_{\uparrow}(x) \; \psi_{\downarrow}(x) \right)^T$ from \ref{eq:eigenstatessup}, the probability distribution on the Bloch sphere is given from
\begin{equation}
\rho(\theta) = \int \mathrm{d} z \left(|\psi_{\uparrow}|^2 + |\psi_{\downarrow}|^2\right) \delta\left(\theta-\pi-2\arctan\left(\frac{\psi_{\uparrow}}{\psi_{\downarrow}}\right)\right),
\end{equation}
The entanglement moments, $C^2_{(n)}(g)$, can be plotted by extracting the Fourier components of the surface and adding them together appropriately.

\end{document}